# Estudiando las estructuras a gran escala del Universo con la teoría de perturbaciones no lineal

Dante V. Gomez-Navarro[1]

[1]Instituto de Física, Universidad Nacional Autónoma de México, Ciudad Universitaria, CP 04510, México D.F., México
E-mail: dantegomezn@gmail.com

*Resumen* — **Las estructuras a gran escala del Universo proporcionan información importante para la cosmología. Las actuales encuestas espectroscópicas, como DESI, medirán millones de galaxias, permitiendo medir la historia de expansión del cosmos y la tasa de crecimiento con gran precisión. En este trabajo, presentamos la descripción teórica del espectro de potencia de materia oscura en el espacio real y en el espacio del corrimiento al rojo para galaxias. Calculamos el espectro usando la teoría de perturbaciones no lineal, la teoría de campo efectiva y el resumado del infrarrojo.**

*Palabras Clave* – **Espectro de potencia, estructuras a gran escala del Universo, teoría de perturbaciones no lineal**

*Abstract* — **The Large-Scale Structure (LSS) of the Universe provides valuable information for modern cosmology. The future redshift surveys, as DESI (Dark Energy Spectroscopic Instrument), will measure millions of galaxies and quasars, which allows to measure the history expansion and the growth factor. In this work, we present the theoretical model of the matter power spectrum and the redshift space multipoles of the galaxy power spectrum. We compute the spectrum using different methods as 1-loop Standard Perturbation Theory, Effective Field Theory and the Infrared resummations.**

*Keywords* — **Galaxy clustering, large-scale structure of the Universe, non-linear perturbation theory**

## I. Introducción

El estado de las observaciones espectroscópicas del Universo permite obtener la medición del crecimiento de las estructuras, fundamental para entender la naturaleza de la energía oscura. Para ello, se usan los catálogos de galaxias y se mide el efecto de las distorsiones del corrimiento al rojo mediante el uso de las estadísticas de dos puntos: el espectro de potencia en el espacio de Fourier, o la función de correlación en el espacio de configuraciones.

Las distorsiones al corrimiento al rojo son una medida de las velocidades peculiares de las galaxias a lo largo de la línea de visión. Las velocidades peculiares son originadas por la interacción gravitatoria entre los objetos y por tanto afecta el crecimiento de las fluctuaciones de materia. En este sentido, cuando se mide el efecto de las distorsiones en el espectro de potencia de las galaxias se puede restringir el factor de crecimiento logarítmico, $f$.

Sin embargo, la tasa de crecimiento está degenerado con el parámetro $\sigma_8$, la amplitud de las fluctuaciones de materia oscura a una escala de $8h^{-1}Mpc$. Por esta razón, el estudio del espectro de potencia en el espacio del redshift es sensible a la combinación de ambos parámetros, lo cual nos referimos simplemente como $f\sigma_8$.

Para entender la riqueza de los datos futuros dentro de los métodos analíticos y semianalíticos, una teoría sólida es necesaria. El universo que observamos contiene estructuras, esencialmente, en todas la escalas. Las estructuras a gran escala es bien modelado por la teoría lineal. Conforme la profundidad y el volumen de los catálogos incrementa, se abarca escalas donde los efectos cuasilineales son más relevantes y la teoría de perturbaciones se convierte más importante.

## II. Espectro de potencias para materia oscura

Los primeros intentos de la descripción de las Estructuras a Gran Escala del universo (LSS) provienen en la década de los 70s [1]. Este esquema es conocido como "Lagrangiano", donde se resuelve perturbativamente el campo de desplazamiento entre la posición inicial y final de las partículas de materia oscura. El esquema más popular ha sido la teoría de perturbaciones estándar (SPT) [2], donde la materia oscura es tratada como un fluido perfecto sin presión y las ecuaciones de movimiento no lineales son resueltas perturbativamente en el espacio Euleriano, donde el campo de densidad de materia $\delta(x,t)$ y el campo de velocidad $\theta(x,t) = \nabla \cdot v(x,t)$ son expandidos en series de Taylor:

$$\delta(x,t) = \delta^{(1)}(x,t) + \delta^{(2)}(x,t) + \delta^{(3)}(x,t) + \cdots$$
$$\theta(x,t) = \theta^{(1)}(x,t) + \theta^{(2)}(x,t) + \theta^{(3)}(x,t) + \cdots \quad (1)$$

A escalas grandes, el campo de densidad de la materia oscura sigue la evolución lineal, es decir, su espectro de potencia está dado por

$$P_{lin}(z,k) = D^2(z)P_{lin}(k) \quad (2)$$

donde $D(z)$ es el factor de crecimiento lineal y $P_{lin}(k)$ es el espectro de potencia lineal al día de hoy ($z = 0$). La primera corrección al espectro de potencias es lo que se conoce como la contribución a un bucle (1-loop, SPT, por sus siglas en inglés), donde los campos de densidad son expandidos

hasta tercer orden. De esta manera, el término a un bucle en la teoría estándar tiene la siguiente estructura

$$P_{1-loop,SPT}(z,k) = D^4(z)\big(P_{13}(k) + P_{22}(k)\big). \quad (3)$$

En la Ec. (3) hemos supuesto un universo Einstein-de-Sitter ($\Omega_m = 1$), en donde las dependencias temporales y de momento se pueden factorizar. En general, las dependencias de momento y del tiempo en las integrales de un bucle no se pueden factorizar. En nuestro caso, los términos $P_{13}(k), P_{22}(k)$ están dados por

$$P_{13}(k) = 6 P_{lin}(k) \int_{\mathbb{q}} F_3(\mathbb{k}, -\mathbb{q}, \mathbb{q}) P_{lin}(q)$$

$$P_{22}(k) = 2 \int_{\mathbb{q}} F_2^2(\mathbb{q}, \mathbb{k}-\mathbb{q}) P_{lin}(q) P_{lin}(|\mathbb{k}-\mathbb{q}|) \quad (4)$$

donde hemos usado la notación $\int_{\mathbb{q}} = \int \frac{d^3 q}{(2\pi)^3}$ y $\mathbb{q}, \mathbb{k}$ representan vectores. Los kernels $F_2, F_3$ son los kernels usuales de la teoría de pertubaciones [2]. Sin embargo, para el entendimiento de las pequeñas escalas, $k_{max} \sim 0.3 h Mpc^{-1}$, es necesario entender la dinámica de las distancias cortas. Note que las integrales en la Ec. (4) son del tipo $I(k) = \int_{\mathbb{q}} M(\mathbb{k}, \mathbb{q})$, las cuales son calculadas sobre todo el espacio de momentos. El formalismo de la teoría de campo efectiva considera una escala de corte para las integrales de un bucle, suavizando el campo de densidad en una escala arbitraria e introduce contratérminos necesarios para remover la dependencia de la escala de corte en las expresiones finales.

En el espacio real, la corrección principal de la teoría efectiva para el espectro de potencias es $-2c_s^2(z)k^2 P_{lin}(z,k)$, donde $c_s$ es la velocidad de sonido de la materia oscura surgido de las ecuaciones de movimiento del fluido para un fluido no perfecto. El parámetro $c_s(z)$ no es conocido a priori y es determinado en el análisis de datos. El espectro de potencia de materia en el esquema de la teoría efectiva (EFT, por sus siglas en inglés) está dado por [3,4]

$$P_{EFT}(z,k) = P_{1-loop,SPT}(z,k) - 2c_s^2(z)k^2 P_{lin}(z,k). \quad (5)$$

La Fig. (1) muestra diferentes modelos para el espectro de potencia de materia oscura. Se considera el efecto del resumado IR, el cual es producida por los desplazamientos de escalas grandes que producen un amortiguamiento no lineal en el espectro de potencia, desfasando las oscilaciones acústicas de bariones (BAO, por sus siglas en inglés). Sin este procedimiento la corrección de EFT no captura la forma de las oscilaciones de BAO e incluso su frecuencia (vea la Sec. V. para una mayor discusión). Este resultado muestra que el resumado IR es un ingrediente necesario para un modelo realístico de las simulaciones de n-cuerpos (puntos en la Fig. (1)). Note que la contribución de EFT mejora la contribución con respecto a la corrección de 1-loop. Por comparación, en línea raya punta color verde se muestra el espectro de potencia lineal.

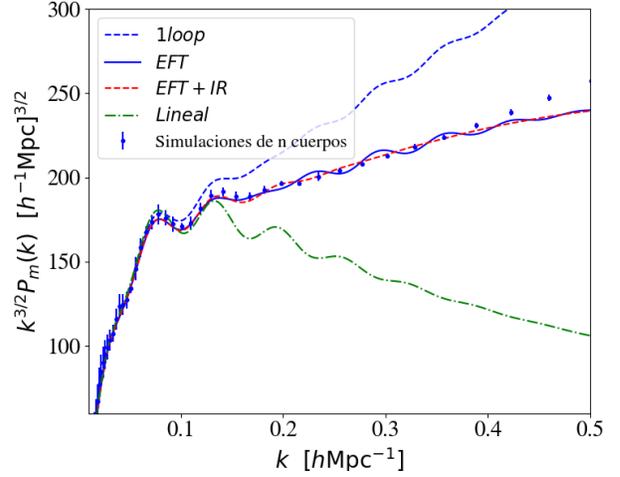

Fig. 1. El espectro de potencia de materia oscura con y sin el resumado IR, junto con la teoría de pertubaciones de 1-loop, y la teoría lineal. Se compara con datos simulados de n cuerpos para ver la precisión de las diferentes técnicas.

### III. ESPECTRO DE POTENCIAS PARA TRAZADORES SESGADOS

Lo que se observa en las encuestas espectroscópicas es el corrimiento el rojo de las galaxias. La forma usual para describir el campo de densidad de cualquier trazador en el Universo, en particular de las galaxias, es haciendo una suma lineal sobre un conjunto finito de operadores ($\delta, \delta^2, \mathcal{G}_2, \delta^3, \delta\mathcal{G}_2, \mathcal{G}_3, \Gamma_3, \partial^2\delta$), donde cada uno de estos operadores tiene asociado un parámetro de sesgo denotado por $b_X$ [5,6,7]. La expansión del sesgo hasta tercer orden está dada por

$$\delta_g = b_1 \delta + \epsilon + \frac{b_2}{2}\delta^2 + b_{\mathcal{G}_2}\mathcal{G}_2 + \frac{b_3}{6}\delta^3 + b_{\delta\mathcal{G}_2}\delta\mathcal{G}_2 + b_{\mathcal{G}_3}\mathcal{G}_3 + b_{\Gamma_3}\Gamma_3 + R^2 \partial^2 \delta, \quad (6)$$

donde hemos definido

$$\mathcal{G}_2(\Phi_g) \equiv (\partial_i \partial_j \Phi_g)^2 - (\partial_i^2 \Phi_g)^2,$$

con $\Phi_g$ el potencial gravitacional. Por otra parte, el operador cúbico que es una contribución no trivial al espectro de potencias a orden de un bucle puede ser escrito como

$$\Gamma_3 \equiv \mathcal{G}_2(\Phi_g) - \mathcal{G}_2(\Phi_v),$$

con $\Phi_v$ el potencial de velocidad. El término $\epsilon$ denota la contribución estocástica. En general, $b_1, b_2, b_3, b_{\delta\mathcal{G}_2}, b_{\mathcal{G}_3}, b_{\Gamma_3}$ y $R^2$ son parámetros libres y son determinados en el análisis de datos.

### IV. ESPECTRO DE POTENCIAS PARA TRAZADORES SESGADOS EN EL ESPACIO DEL CORRIMIENTO AL ROJO

En las encuestas de galaxias espectroscópicas, las posiciones son típicamente inferidas midiendo el corrimiento al rojo. El mapeo del espacio del corrimiento al rojo es caracterizado por el coseno del ángulo entre la línea de visión $\mathbb{z}$ y el vector de onda de un modo dado $\mathbb{k}$,

$\mu = (\mathbb{z} \cdot \mathbb{k})/|\mathbb{k}|$, es decir $P_{gg,RSD}(\mathbb{k}) = P_{gg,RSD}(k,\mu)$. El espectro de potencias en el espacio del corrimiento al rojo tiene la siguiente estructura

$$P_{gg,RSD}(z,k,\mu) = Z_1^2(\mathbb{k})P_{lin}(z,k)$$
$$+2\int_q Z_2^2(\mathbb{q},\mathbb{k}-\mathbb{q})P_{lin}(z,|\mathbb{k}-\mathbb{q}|)P_{lin}(z,q)$$
$$+6Z_1(\mathbb{k})P_{lin}(z,k)\int_q Z_3(\mathbb{q},-\mathbb{q},\mathbb{k})P_{lin}(z,q)$$
$$+P_{ctr,RSD}(z,k,\mu) + P_{\epsilon\epsilon,RSD}(z,k,\mu). \quad (7)$$

Las funciones $Z_1(\mathbb{k}), Z_2(\mathbb{k}_1,\mathbb{k}_2), Z_3(\mathbb{k}_1,\mathbb{k}_2,\mathbb{k}_3)$ son los kernels en el espacio del corrimiento que dependen de los parámetros de sesgo, la tasa de crecimiento logarítmica $f$, y de $\mu$ [8].

Los últimos dos términos de la Ec. (7) hacen referencia a las contribuciones de EFT y a la contribución estocástica, respectivamente. Los contratérminos de EFT vienen dados por

$$P_{ctr,RSD}(z,k,\mu) = P_{ctr,RSD,\nabla^2\delta}(z,k,\mu)$$
$$+P_{ctr,RSD,\nabla_z^4\delta}(z,k,\mu). \quad (8)$$

En el espacio del corrimiento al rojo, la situación es más compleja ya que se considera tanto la física de pequeña escala como el mapeo no lineal entre los campos de densidad del espacio del corrimiento al rojo y del espacio real. Cada multipolo del espectro de potencias tiene asociado su contratérmino de la forma $\sum \overline{\alpha}_n(z)\mu^{2n}k^2 P_{lin}(k)$. En este caso, el primer término de la Ec. (8) puede ser visto como una generalización de la velocidad de sonido de la materia oscura $c_s(z)$, dando origen a la siguiente expresión

$$P_{ctr,RSD,\nabla^2\delta}(z,k,\mu) = -2\overline{c}_0(z)k^2 P_{lin}(z,k)$$
$$-2\overline{c}_2(z)f(z)\mu^2 k^2 P_{lin}(z,k)$$
$$-2\overline{c}_4(z)f^2(z)\mu^4 k^2 P_{lin}(z,k) \quad (9)$$

con $\overline{c}_0(z), \overline{c}_2(z)$ y $\overline{c}_4(z)$ son parámetros en unidades de $[h^{-1}Mpc]^2$.

Otro efecto en el espacio del corrimiento al rojo está relacionado con el efecto de los dedos de Dios, acoplamiento no lineal entre los campos de velocidad y de densidad, visible a lo largo de la línea de visión. La escala característica de este efecto está dada por la dispersión de velocidad $\sigma_v$, típicamente $\sigma_v \sim k_{NL}$, dando origen a la siguiente estructura $\overline{\alpha}(z)(\mu k f(z)\sigma_v(z))^4 P_s^K(k,\mu)$. La dispersión de velocidad no es bien modelada por la teoría de perturbaciones. Por esta razón, usualmente se elige a $\overline{c}_{\nabla_z^4\delta}(z) = \overline{\alpha}(z)\sigma_v(z)^4$ como parámetro libre. En resumen, el segundo término de la Ec. (8) hace referencia a la presencia de los dedos de Dios (fingers-of-God), los cuales son inducidos por términos de derivadas superiores, dando origen a la siguiente expresión

$$P_{ctr,RSD,\nabla_z^4\delta}(z,k,\mu) = \overline{c}_{\nabla_z^4\delta}(z)f^4(z)\mu^4 k^4$$
$$\times P_{K,RSD}(k,\mu). \quad (10)$$

El espectro de potencia a orden lineal en el espacio del corrimiento al rojo está dado por $P_{K,RSD}(k,\mu) = (b_1(z) + f(z)\mu^2)^2 P_{lin}(z,k)$, y se le conoce como el espectro de potencia de Kaiser.

Finalmente, el espectro de potencia estocástico lo consideramos como un término constante conocido como Shot-noise, es decir, $P_{\epsilon\epsilon,RSD}(z,k,\mu) = P_{shot}(z) = \frac{1}{\overline{n}}$, donde $\overline{n}$ indica la densidad promedio de objetos en la muestra. En general, $P_{\epsilon\epsilon,RSD}(z,k,\mu)$ tiene dos términos adicionales que dependen de la escala para el monopolo y cuadrupolo. Para nuestros fines, basta solo considerar el término Shot-noise.

Sin embargo, es más conveniente escribir la información angular en multipolos. Los multipolos del espectro de potencia de galaxias están dados por

$$P_\ell(z,k) = A_\ell \int_{-1}^{1} d\mu \mathcal{L}_\ell(\mu) P_{gg,RSD}(z,k,\mu) \quad (11)$$

con $A_\ell = (2\ell+1)/2$.

Las expresiones finales para los multipolos del espectro de potencia de las galaxias pueden ser escritas como [8]

$$P_0(z,k) = \left(P_{0,\theta\theta}^{lin}(z,k) + P_{0,\theta\theta}^{1-loop,SPT}(z,k)\right)$$
$$+b_1(z)\left(P_{0,\theta\delta}^{lin}(z,k) + P_{0,\theta\delta}^{1-loop,SPT}(z,k)\right)$$
$$+b_1^2(z)\left(P_{0,\delta\delta}^{lin}(z,k) + P_{0,\delta\delta}^{1-loop,SPT}(z,k)\right)$$
$$+0.25 b_2^2(z) I_{0,\delta^2\delta^2}(z,k) + b_1(z)b_2(z) I_{0,\delta\delta^2}(z,k)$$
$$+b_2(z) I_{0,\theta\delta^2}(z,k) + b_1(z)b_{\mathcal{G}_2}(z) I_{0,\delta\mathcal{G}_2}(z,k)$$
$$+b_{\mathcal{G}_2}(z) I_{0,\theta\mathcal{G}_2}(z,k) + b_2(z)b_{\mathcal{G}_2}(z) I_{\delta^2\mathcal{G}_2}(z,k)$$
$$+b_{\mathcal{G}_2}^2(z) I_{\mathcal{G}_2\mathcal{G}_2}(z,k)$$
$$+\left(2b_{\mathcal{G}_2}(z) + 0.8 b_{\Gamma_3}(z)\right)\left(b_1(z) F_{0,\delta\mathcal{G}_2}(k) + F_{0,\theta\mathcal{G}_2}(z,k)\right)$$
$$+c_0(z) P_{0,\nabla^2\delta}(z,k) + \overline{c}_{\nabla_z^4\delta}(z) P_{0,\nabla_z^4\delta}(z,k)$$
$$+P_{shot}(z), \quad (12\text{ a})$$

$$P_2(z,k) = \left(P_{2,\theta\theta}^{lin}(z,k) + P_{2,\theta\theta}^{1-loop,SPT}(z,k)\right)$$
$$+b_1(z)\left(P_{2,\theta\delta}^{lin}(z,k) + P_{2,\theta\delta}^{1-loop,SPT}(z,k)\right)$$
$$+b_1^2(z) P_{2,\delta\delta}^{1-loop,SPT}(z,k) + b_1(z)b_2(z) I_{2,\delta\delta^2}(z,k)$$
$$+b_2(z) I_{2,\theta\delta^2}(z,k)$$
$$+b_1(z)b_{\mathcal{G}_2}(z) I_{2,\delta\mathcal{G}_2}(z,k) + b_{\mathcal{G}_2}(z) I_{2,\theta\mathcal{G}_2}(z,k)$$
$$+\left(2b_{\mathcal{G}_2}(z) + 0.8 b_{\Gamma_3}(z)\right) F_{2,\theta\mathcal{G}_2(z,k)}$$
$$+c_2(z) P_{2,\nabla^2\delta}(z,k) + \overline{c}_{\nabla_z^4\delta}(z) P_{2,\nabla_z^4\delta}(z,k), \quad (12\text{ b})$$

$$P_4(z,k) = \left(P_{4,\theta\theta}^{lin}(z,k) + P_{4,\theta\theta}^{1-loop,SPT}(z,k)\right)$$

$$+b_1(z)\left(P^{lin}_{4,\theta\delta}(z,k)+b_1(z)P^{1-loop,SPT}_{4,\delta\delta}(z,k)\right)$$
$$+b_2(z)I_{4,\theta\delta^2}(z,k)+b_{\mathcal{G}_2}(z)I_{4,\theta\mathcal{G}_2}(z,k)$$
$$+c_4(z)P_{4,\nabla^2\delta}(z,k)+\overline{c}_{\nabla_z^4\delta}(z)P_{4,\nabla_z^4\delta}(z,k). \quad (12\text{ c})$$

$P_{\delta\delta}, P_{\delta\theta}, P_{\theta\theta}$ son el espectro del campo de densidad, el espectro cruzado entre el campo de densidad y velocidad, y el espectro del campo de velocidad, respectivamente. Las expresiones del tipo $I_X(z,k), F_X(z,k)$ están definidas en [8]. Por otra parte, los nuevos coeficientes $c_0, c_2, c_4$ están definidos como

$$c_0 \equiv \overline{c}_0 + \frac{f}{3}\overline{c}_2 + \frac{f^2}{5}\overline{c}_4,$$
$$c_2 \equiv \overline{c}_2 + \frac{6f}{7}\overline{c}_4, \quad c_4 \equiv \overline{c}_4 \quad (13)$$

La relación (13) no es exacta cuando se considera la técnica del resumado del infrarrojo que veremos en la siguiente sección.

## V. Resumado del infrarrojo (IR)

Los desplazamientos de escalas grandes producen un amortiguamiento no lineal con huellas localizadas en el espectro de potencia tal como las oscilaciones acústicas de bariones (BAO). Si bien la dinámica de estas estructuras es esencialmente lineal, el tamaño de los desplazamientos en las escalas de BAO pueden ser del mismo orden al número de onda donde las oscilaciones de BAO residen. Los efectos de estos desplazamientos pueden ser manualmente resumados en las expansiones tal como se hace en la teoría de perturbaciones. La técnica que permite identificar y resumar las contribuciones del infrarrojo que alteran el patrón de BAO se le conoce como resumado del infrarrojo (IR) [9]. Puesto que el flujo de los bultos grandes afecta las escalas de BAO, el punto de partida inicia con la división del espectro de potencia lineal en la componente suavizada $P_{nw}(k)$ y en la parte oscilatoria $P_w(k)$, de tal forma que
$$P_{lin}(k) = P_{nw}(k) + P_w(k).$$

El espectro de potencia resumado se convierte en
$$P^{IR}_{gg}(z,k,\mu) = (b_1(z) + f(z)\mu^2)^2$$
$$\times\left(P_{nw}(z,k) + e^{-k^2\Sigma^2_{tot}(z,\mu)}P_w(z,k)\left(1 + k^2\Sigma^2_{tot}(z,\mu)\right)\right)$$
$$+P_{gg,nw,RSD,EFT}(z,k,\mu)$$
$$+e^{-k^2\Sigma^2_{tot}(z,\mu)}P_{gg,w,RSD,EFT}(z,k,\mu) \quad (14)$$

donde la contribución de la teoría efectiva total está dada por
$$P_{gg,RSD,EFT}(z,k,\mu) = P_{gg,RSD,1-loop,SPT}(z,k,\mu)$$
$$+P_{ctr,RSD}(z,k,\mu).$$

La función de amortiguamiento total está definida como
$$\Sigma^2_{tot}(z,\mu) = (1+f(z)\mu^2)\left((2+f(z))\right)\Sigma^2(z)$$
$$+f^2(z)\mu^2(\mu^2-1)\delta\Sigma^2(z),$$

donde $\Sigma^2(z)$ es la función de amortiguamiento en el espacio real
$$\Sigma^2(z) \equiv \frac{1}{6\pi^2}\int_0^{k_s}dq\,P_{nw}(z,q)$$
$$\times\left[1-j_0\left(\frac{q}{k_{osc}}\right)+2j_2\left(\frac{q}{k_{osc}}\right)\right].$$

El número de onda correspondiente a la longitud de onda de BAO ($\ell_{osc} 110 hMpc^{-1}$) es $k_{osc}$, y la nueva contribución en el espacio al corrimiento al rojo es
$$\delta\Sigma^2(z) \equiv \frac{1}{2\pi^2}\int_0^{k_s}dq\,P_{nw}(z,q)j_2\left(\frac{q}{k_{osc}}\right).$$

Los términos $P_{...EFT}[P_{lin}]$ son tratados como funcionales del espectro de potencia lineal, es decir,
$$P_{gg,nw,RSD,EFT}(z,k,\mu) = P_{gg,RSD,EFT}[P_{nw}],$$
$$P_{gg,w,RSD,EFT}(z,k,\mu) = P_{gg,RSD,EFT}[P_{nw}+P_w]$$
$$-P_{gg,RSD,EFT}[P_{nw}].$$

En la Fig. (2) se muestra las predicciones teóricas incluyendo el resumado IR y sin el. El modelo es consistente con los valores extraídos de los catálogos de galaxias de eBOSS (vea [10] para una mayor discusión).

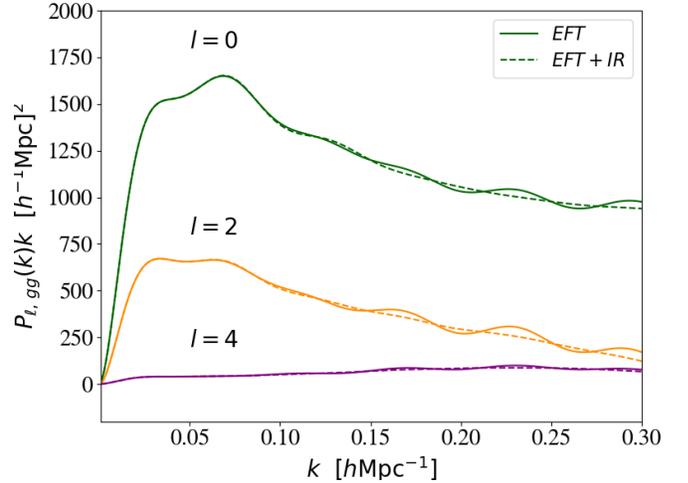

Fig. 2. Los multipolos en el espacio del corrimiento al rojo del espectro de potencia de galaxias con y sin el resumado IR.

## IV. Discusión

Vivimos en la era de oro de la cosmología donde las próximas encuestas de galaxias espectroscópicas proporcionarán información sobre la tasa de crecimiento de estructuras y de expansión con una precisión porcentual. Encuestas de galaxias espectroscópicas, como DESI, ya han iniciado esta aventura siendo capaces de crear el mapa 3D más grande del Universo a finales del año 2025. Para describir estos grandes catálogos es necesario construir modelos del espectro de potencia en el espacio del corrimiento al rojo.

## V. Conclusión

Con base en la teoría de perturbaciones estándar, la teoría de campo efectiva y el esquema del resumado del infrarrojo se construyó los multipolos del espectro de potencia de las galaxias. Los contratérminos de la teoría efectiva son necesarios para modelar la física de pequeña escala que está fuera del alcance de la teoría de perturbaciones estándar. Mientras que el resumado del infrarrojo sirve para modelar los flujos de los grandes bultos que afectan las oscilaciones de BAO. Estas técnicas permiten estudiar los datos observados por el catalógo de galaxias eBOSS (extended Baryon Oscillation Spectroscopic Survey) [10].